\documentclass[apjl]{emulateapj}
\usepackage{apjfonts}
\submitted{Received 2007 October 5; accepted 2007 November 8}
\journalinfo{\scshape The Astrophysical Journal, Preprint}

\begin{document}
\title{THE SPATIAL DISTRIBUTION OF HARD X-RAY SPECTRAL INDEX AND LOCAL MAGNETIC\\RECONNECTION RATE}
\author{\normalsize\scshape Chang Liu,\altaffilmark{1} Jeongwoo Lee,\altaffilmark{2} Ju Jing,\altaffilmark{2} Dale E. Gary,\altaffilmark{2} And Haimin Wang\altaffilmark{1,2}}
\altaffiltext{1}{Big Bear Solar Observatory, New Jersey Institute of Technology, 40386 North Shore Lane, Big Bear City, CA 92314-9672; cliu@bbso.njit.edu}
\altaffiltext{2}{Center for Solar-Terrestrial Research, New Jersey Institute of Technology, University Heights, Newark, NJ 07102-1982}

\begin{abstract}
The rare phenomenon of ribbon-like hard X-ray (HXR) sources up to 100~keV found in the 2005 May 13 M8.0 flare observed with the \textit{Reuven Ramaty High Energy Solar Spectroscopic Imager} provides detailed information on the spatial distribution of flare HXR emission. In this Letter, we further investigate the characteristics of HXR emission in this event using imaging spectroscopy, from which we obtain spatially resolved HXR spectral maps during the flare impulsive phase. As a result we found, along a flare ribbon, an anticorrelation relationship between the local HXR flux and the local HXR spectral index. We suggest that this can be regarded as a spatial analog of the well-known temporal soft-hard-soft spectral evolution pattern of the integrated HXR flux. We also found an anticorrelation between HXR spectral index and local electric field along the ribbon, which suggests the electron acceleration by the electric field during flares.
\end{abstract}


\section{INTRODUCTION}
It has long been recognized that hard X-ray (HXR) emission is a powerful diagnostic of accelerated energetic electrons produced by flares. The HXR spectrum emitted by nonthermal electrons often appears as a power-law distribution in photon energy, which implies a characteristic in the energy distribution of the electron flux bombarding the target under the bremsstrahlung emission mechanism. Temporal evolution of the electron energy distribution is highly important for identifying the dominant acceleration process.

As known from the early results of the traditional scintillation-counter spectrometers, the HXR spectral index generally follows a soft-hard-soft (SHS) spectral pattern in the rise-maximum-decay phase of flares \citep{parks69,benz77,brown85,dennis85}. It was further corroborated by the results of \textit{Reuven Ramaty High Energy Solar Spectroscopic Imager} \citep[\textit{RHESSI};][]{lin02} that can measure HXR spectra with much higher resolution. \citet{hudson02} illustrated the consistency of the SHS pattern derived using higher energy resolution \textit{RHESSI} data (of order 1~keV) with that obtained with the HXRS scintillation-counter spectra. A systematic study by \citet{grigis04} using \textit{RHESSI} revealed that the SHS pattern appears even in sub-peaks of HXR with durations of one minute to shorter than 8~s. With \textit{RHESSI}'s capability of imaging spectroscopy, the study of SHS has been extended to spatially resolved HXR sources. \citet{emslie03} and \citet{battaglia06} examined spectral evolution of both coronal and footpoint HXR sources in the course of flares and find that they commonly show the SHS behavior. Based on the results they suggested that SHS can be an intrinsic feature of the electron acceleration in the flare impulsive phase.

The SHS can also be regarded as a relation between HXR spectral index and flux because the hardest spectrum usually appears in the period of maximum flux \citep{grigis06}. As a possible extension of this relation to spatial characteristics, we may consider a spatial analog of the SHS pattern in which a stronger HXR region has a harder spectrum. Such idea has been proposed in a couple of studies. For instance, \citet{masuda01} determined spectral index maps of an extended HXR event using \textit{Yohkoh} data to find spectral hardening toward the ribbon edge where the most intense energy release is expected. \citet{hudson04} proposed that the regions of weaker HXR emission would correspond to softer HXR spectra, in an attempt to explain the footpoint-like HXR morphology in contrast with extensive ribbon structure seen at H$\alpha$ and UV wavelengths. However, it will need a more thorough examination to see whether the relation between HXR flux and spectral index in time can simply be transformed to such a spatial analog. This is one of the goals of this paper that we intend to achieve using \textit{RHESSI} imaging spectroscopy.

In addition to spectral evolution, the HXR flare ribbon motion can also manifest the progression of flare energy release due to magnetic reconnection. This was enabled after \citet{forbes84} derived a relationship between the ribbon expansion velocity $u$ and the electric field $E$ in the reconnecting current sheet (RCS) in the form of $E = uB$, where $B$ is the local vertical magnetic field strength in the footpoint. It is important to have an indirect measurement of the electric field, because direct acceleration of electrons by the electric field is a candidate mechanism for creating high energy electrons in flares \citep{litvinenko96}. Some studies showed that there is a temporal correlation between HXR flux and the electric field derived this way \citep[e.g.,][]{qiu02}. It is then of next interest whether there is not only a temporal but also a spatial correlation between HXR spectral index and electric field in the RCS. Recently, \citet{wood05} presented a particle simulation with prescribed electric and magnetic fields, which shows a hardening of the electron energy spectrum with increasing electric field. Such a model prediction can be compared with the above mentioned measurements of the HXR spectrum and electric field.

In this Letter, we investigate spatial distribution of HXR spectrum and electric field during the 2005 May 13 M8.0 flare. This is a particular event \citep{liu07c}, where the HXR sources are so extended as to be suitable for the imaging spectroscopy, and is thus adequate for studying the possible spatial relationship between the spectral index and electric field.

\section{IMAGING SPECTROSCOPY}

\begin{figure}[t]
\epsscale{1.15}
\plotone{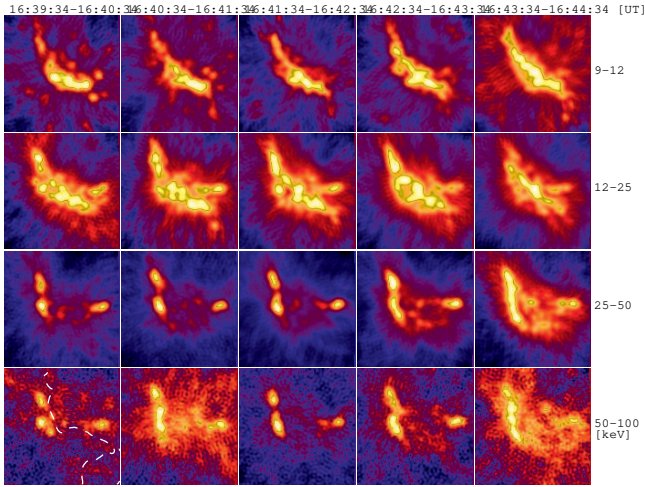}
\caption{\textit{RHESSI} CLEAN images for five one-minute time intervals in four energy bands from 9--100~keV. The images are centered on the point (-168\arcsec, 230\arcsec) with solar west to the right and north up. The field of view is 128\arcsec $\times$ 128\arcsec. The white dashed line superposed on the lower left panel denotes the magnetic polarity inversion line. \label{f1}}
\end{figure}

\textit{RHESSI} imaging spectroscopy has thus far been made in such a way to determine the flux spectra integrated over spatially isolated sources \citep[e.g.,][]{emslie03,battaglia06}. This technique gives a viable result when the source is so compact as to be barely resolved. In our case, we however study extended HXR sources and our goal is to explore the spatial variation of spectral index inside the sources. We thus choose to measure local HXR spectral index based on spectrum in each pixel. Previously, such attempt was also made for the only other reported ribbon-like HXR event, in which \citet{masuda01} obtained spectral maps providing the spectral index information in each pixel, from the count ratio between \textit{Yohkoh} M2- and H-band. With \textit{RHESSI}'s higher temporal and energy resolution as well as improved dynamic range, we expect a better imaging spectroscopic result. For this moderate event, however, we still suffer from the trade off between choosing finer energy/time bins and making better images by accumulating for wider energy/time bins.

Most often, \textit{RHESSI} imaging is made using the CLEAN algorithm with grids 3--9. Recently, \citet{brian07} showed that adding grids 1 and 2 (the finest grids) can enhance the image quality depending on the source structure, and the ribbon-like HXR sources in this event is such a suitable case. We found that use of all grids not only improves the imaging quality compared with using grids 3--9, but also gives better count statistics that is essential for imaging spectroscopy. We also found that natural weighting produces a more physically reasonable source morphology for this event \citep{liu07a}. While PIXON or NJIT-MEM may give more details in the source morphology, CLEAN gives a reasonable result for imaging spectroscopy \citep{battaglia06}. We thus use the CLEAN algorithm along with all grids and natural weighting, which gives a FWHM resolution of $\sim$5.9\arcsec.

We first make 128~$\times$~128 pixels images with a pixel size of 1\arcsec\ for the five one-minute intervals that cover the rise and decay phases of HXR \cite[$a$--$e$ as in][Fig.~1]{liu07a}. Figure~\ref{f1} shows the images in four energy bands from 9--100~keV. The emission at high energies ($\geq$25~keV) are mainly sources located in each side of the magnetic polarity inversion line (PIL) presumably from the footpoints of the flaring loops. A footpoint-to-ribbon transformation of the HXR morphology can be clearly seen across the peak of the HXRs at 16:42:04~UT. At lower energies, sources above the PIL become evident and they could come from the tops of the loops joining the HXR footpoints and ribbons, while the footpoint emissions are still visible, as low as in 9--12~keV band. We emphasize that the ribbon-like HXR sources are an intrinsic feature in this event, as evidenced by its fidelity to the UV ribbon emissions \citep{liu07a,brian07}.

\begin{figure}[t]
\epsscale{1.15}
\plotone{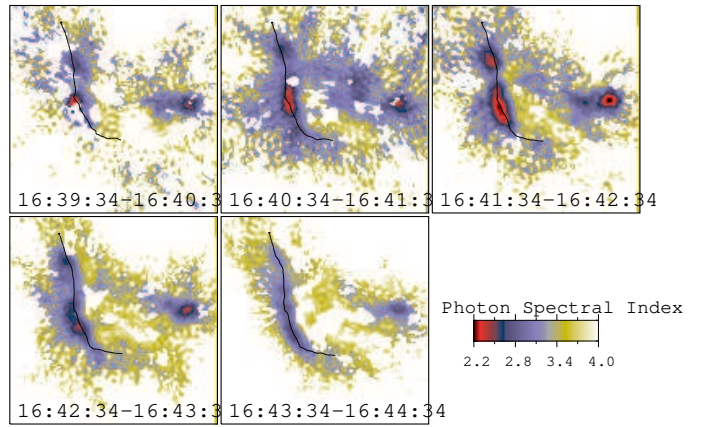}
\caption{HXR spectral maps in the five one-minute time intervals in Fig.~\ref{f1} covering the HXR rise to decay phases. These maps are generated based on HXR images in Fig.~\ref{f1} assuming a single power-law spectrum at each pixel. A lower power-law index corresponds to a flatter spectrum. The white background indicates the position where no good power-law fitting can be obtained. The black curves are our trace-outs of multiple locations along H$\alpha$ ribbons determined by finding out local maximum H$\alpha$ intensity in each time interval. The curves are composed of 100 points evenly distributed along the ribbon. \label{f2}}
\end{figure}

We then derive spectral index maps from the four energy bands shown in Figure~\ref{f1}, assuming that the local HXR spectrum in each pixel follows a power-law distribution in photon energy with spectral index defined by $\gamma = -d {\rm log} I/d{\rm log} \epsilon$. While it is more desirable to have narrower energy bands for a better spectroscopy, the choice of only four energy band is inevitable to ensure enough photon statistics in each time interval for this event. Another issue in HXR imaging spectroscopy is that the coronal and footpoint sources show different spectral characteristics. Coronal HXR sources usually show thermal-like spectrum and footpoint sources, power-law spectrum \citep{battaglia06}, and we had better distinguish them from each other. In the present event, the HXR emission is dominated by the footpoint sources in a wide energy range $\sim$10--100~keV and we fit all sources in the energy range to power-law spectra. Nevertheless, our approximation that each photon spectrum measured in four energy bands should follow a single power law may not be good enough in all pixels. We thus set a threshold, based on the chi-square of a value of 0.25 returned by the LINFIT procedure of IDL, for the goodness of the fitting, below which we do not calculate the spectral index. The results for each time interval are shown in Figure~\ref{f2}, where the color coded spectral index is displayed only for those fittable pixels. To ascertain that the flux in each pixel is enough for generating a reasonable spectrum, we further reduce the images to 32~$\times$~32 pixels by averaging a set of 4~$\times$~4 pixels into one. We repeat the fitting procedure and obtain nearly identical spectral features.

\section{RESULTS} \label{res}

\begin{figure*}[t]
\epsscale{1.15}
\plotone{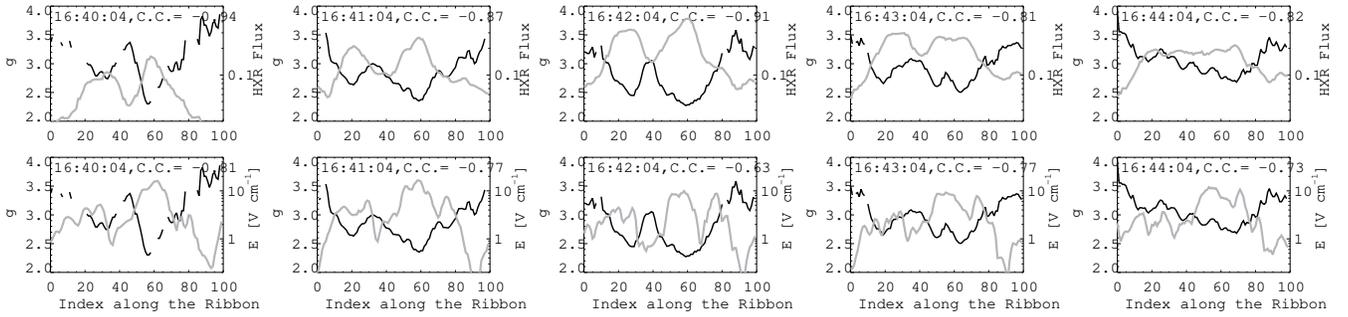}
\caption{Spatial distribution of HXR spectral index (\textit{black}) in comparison with those of HXR flux in the 25--100~keV range (\textit{grey, upper panels}) and electric field (\textit{grey}, lower panels) in the five one-minute time intervals, shown as functions of the ribbon distance. Depicted as the black line in Fig.~\ref{f2}, the index of the ribbon distance runs from 0 in the northern end to 99 in the southern end. The existence of break points in the profiles of spectral index indicates bad fittings at those locations. \label{f3}}
\end{figure*}

Comparison of Figure~\ref{f1} and Figure~\ref{f2} reveals several characteristics. First, there are three major footpoint-like HXR emitting sources with hard spectra near the HXR peak at $\sim$16:42:04~UT, when the spectral index reaches a minimum value of $\sim$2.2. The averaged value of spectral index over the fittable pixels in the field of view is $\sim$3.3 at this time, which agrees with that derived using the OSPEX package for integrated X-ray emission. Earlier and later in the event, the overall spectrum of the flare region as well as those of the main HXR sources are seen to be much softer, with higher index values. This temporal evolution of the SHS pattern is thus what typically has been observed before \citep{battaglia06}. Second, the main HXR sources show a spatial distribution of spectral index from the center with smaller value (harder spectrum) to the outer regions with larger value (steeper spectrum), which is most prominent near the HXR peak (see panel 16:41:34--16:42:34~UT). When the HXR evolve to a ribbon morphology later in the event (e.g., panel 16:43:34--16:44:34~UT), there are still kernels with harder spectra discernible, although the whole system has a much steeper spectrum compared with the flare peak. 

As our major interest lies in how the physical quantities vary spatially, we make the comparison between the HXR spectral index and flux along the ribbon axis. Specifically, we measure the variation of spectral index along the eastern flare ribbon, where there is a clear footpoint-to-ribbon evolution of HXR morphology \citep[c.f. Fig.~\ref{f1}; also see][]{liu07a}. We trace out the spectral index and the flux in the 25--100~keV range using the same indexes of ribbon distance as defined by \citet[][specified in Fig. 2 as black line at each time interval]{jing07}, and present the results in the upper panels of Figure~\ref{f3}. It is obvious in each time interval that the HXR spectral index exhibits a strong spatial anticorrelation with the HXR flux ($\log F$), with absolute values of the correlation coefficient $\gtrsim$0.8. In the lower panels of Figure~\ref{f3}, we compare the spatial evolution of the HXR spectral index with that of the electric field ($\log E$) in the RCS, which was previously derived by tracing the H$\alpha$ ribbon motion and incorporating the longitudinal magnetic field measurement \citep{jing07}. It can be seen that there also exists a prominent anticorrelation relationship between these two quantities, with absolute values of the correlation coefficient $\gtrsim$0.65.

\begin{figure}[t]
\epsscale{1.}
\plotone{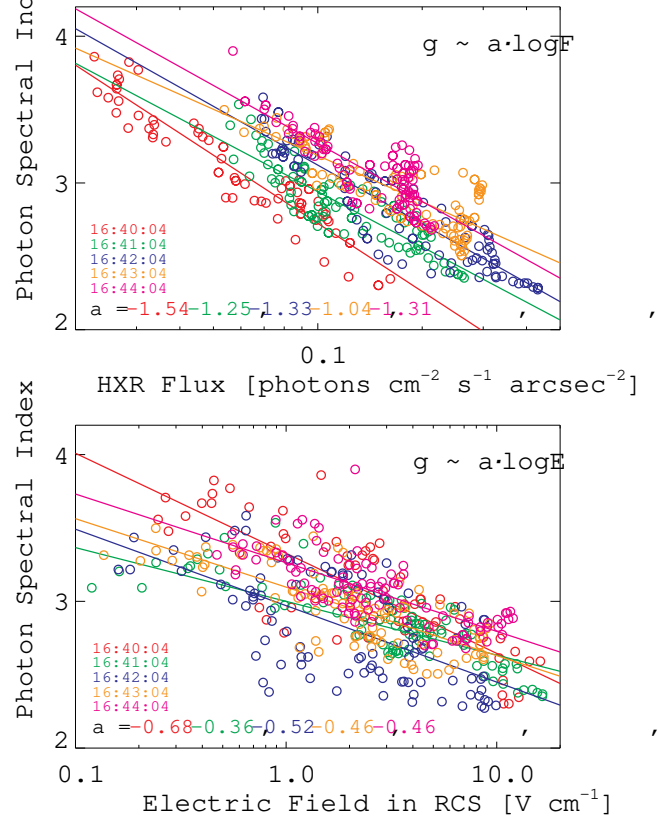}
\caption{Scatter plot of spectral index ($\gamma$) and HXR flux ($\log F$, upper panel), and spectral index and electric field ($\log E$, lower panel), measured at each indexed location along the ribbon for all the time intervals. Data points and the best fits in linear-log space (solid lines) by minimizing the chi-square error statistic for different time intervals are represented with different colors. \label{f4}}
\end{figure}

Figure~\ref{f4} shows the results in Figure~\ref{f3} as scatter plots. The relationship between the spectral index and flux and the spectral index and electric field are shown in the upper and lower panel, respectively. First, we can see without ambiguity that at a specific time, the source position with a weaker HXR emission (lower HXR flux) correspond to steeper X-ray spectra and presumably to softer electron precipitation spectra. We therefore suggest that this anticorrelation between HXR spectral index and flux is a spatial analog of the well-known temporal SHS pattern of HXR emission. Second, the anticorrelation relationship between the spectral index and electric field clearly points to a softer HXR spectrum in the case of weaker electric field, consistent with the trend predicted by numerical simulation \citep{wood05}. This result thus appears to support the hypothesis that direct acceleration by the electric field in the RCS may play an important role in producing energetic electrons in flares.

\section{SUMMARY AND DISCUSSION}

Exploiting \textit{RHESSI}'s capability of imaging spectroscopy, we have determined spectral index of the local photon spectrum in the unusually extended HXR source observed in the 2005 May 13 M8.0 flare. We then find a spatial anticorrelation relationship between both the local HXR flux and electric field corresponding to the local HXR spectral index at several time intervals. We discuss the present results in comparison with similar works on solar HXR imaging spectroscopy.

The present approach is closest to that of \citet{masuda01} who used \textit{Yohkoh} images at two energy bands to find a spectral index change across flare ribbons. In their result, the HXR spectrum at the outer edge of the ribbon is found to be harder than in other regions of the ribbon and the hardness of the spectrum gradually changes across the ribbon width. This is consistent with the physical picture that the ribbon edge is connected to the most recently reconnected field lines and thus shows the most energetic electrons. The present study shows a similar result with more spatial details. It clearly shows the hardest HXR sources lying along the edge of flare ribbon in UV \cite[c.f. Fig~\ref{f2} and][Fig.~2]{liu07a} and a smooth transition of spectral index across the HXR sources. Particularly, we find that the HXR spectral index exhibits a strong spatial anticorrelation with the HXR flux in all time intervals during the flare impulsive phase. We call this \textit{spatial SHS} behavior in analogy with the well-known temporal SHS pattern of integrated HXR flux. The spatial SHS may also be an essential feature of solar flare electron acceleration, and may help explaining the confined nature of HXR sources compared with extended H$\alpha$ and UV ribbons \citep[c.f.][]{hudson04}.

We must note that the spatial SHS implies a more strict relationship between HXR flux and spectral index than that found in the previous \textit{RHESSI} studies on the temporal SHS behavior in individual sources \citep{emslie03,battaglia06}. In the latter results, each isolated source exhibits the temporal SHS pattern independent of each other and thus the normalization of the flux--spectral index relation ($F$--$\delta$ relation) may differ from one source to another. On the other hand, our spatial SHS implies that the same normalization of the $F$--$\delta$ relation should hold in all regions. A major difference between those works and ours lies in that our spatial SHS refers to correlations among local sub-regions within one footpoint side of the magnetic arcade, whereas \citet{emslie03} and \citet{battaglia06} compared spectra integrated over individual footpoint sources. We suspect that almost all local regions in this event were subject to a common acceleration and transport process to share the same $F$--$\delta$ relation. For this reason we do not believe that our result is in conflict with the previous results.

An entirely new result in this study is the spatial anticorrelation between HXR spectral index and electric field in the RCS. This property was found because we could measure both the spectral index and the electric field as functions of position within an extended HXR ribbon. As a comparison, we note that the numerical simulation for the direct electric field acceleration of electrons predicted electron energy distribution with power law index $\delta=4.0, 2.6, 1.5$ for electric field strength $E =$~0.1, 1.0, 10~V~cm$^{-1}$, respectively \citep{wood05}. We are not sure how to convert this electron power law index $\delta$ into the observed photon spectral index $\gamma$, because further assumptions need to be made on the nature of the numerical solutions and the radiation. Within the scope of this paper, we note that there is a qualitative agreement between our observational result (Fig.~\ref{f4}) and the model result \citep{wood05}, in that both indicate a hardening of electron energy distribution with increasing electric field strength. It appears that the present result suggests the dominance of direct electric field acceleration of flare electrons. It however does not exclude other possibilities. As discussed by \citet{hudson02}, an explicit theoretical demonstration of the SHS behavior under the framework of the thick-target model can be made with the stochastic acceleration mechanism discussed by \citet{benz77}. Alternatively it is still possible that the efficient electron acceleration is confined in the regions of strong HXR flux and the weaker HXR flux regions nearby result from some propagation effect. A more systematic survey of the relationship between these two physical quantities, both temporally and spatially, will be needed in order to ascertain such association under the context of electron acceleration mechanism.

\acknowledgments
The authors thank the \textit{RHESSI} team for the excellent data set. We also thank the referee for his/her helpful comments that improved the Letter. C. L. is indebted to A.~G. Emslie, B.~R. Dennis, and R. A. Schwartz for valuable discussions. C. L. also thanks Y. Xu and H. Ji for help. C. L., J. J., and H. W. were supported by NSF/SHINE grant ATM 05-48952 and NSF grant ATM 05-36921 and NASA grant NNX0-7AH78G. J. L. and D. E. G. were supported by NSF grant AST 06-07544 and NASA grant NNG0-6GE76G.

\end{document}